# Skew scattering dominated anomalous Nernst effect in monovalent cation doped bulk La$_{1-x}$Na$_x$MnO$_3$

Arup Ghosh, Rajasree Das, and Ramanathan Mahendiran*

*Department of Physics, 2 Science Drive 3, National University of Singapore, Singapore 117551, Republic of Singapore*

## ABSTRACT

We report the anomalous Nernst effect (ANE), i.e. creation of electrical voltage transverse to the direction of applied temperature gradient and magnetic field in bulk samples of La$_{1-x}$Na$_x$MnO$_3$ ($x$ = 0.02, 0.05, 0.1 and 0.3) whose ferromagnetic Cure temperature($T_C$) varies from 251 K for $x$ = 0.02 to 310 K for $x$ = 0.3. We measured both the magnetic field and temperature dependences of ANE. While ANE is negligible much above $T_C$, it increases rapidly at $T_C$ and exhibits a peak just below it. The field dependence of ANE closely follows magnetization. The maximum value of ANE at $T_C$ increases from $x$ = 0.02 to 0.1, but decreases significantly for $x$ = 0.3. Besides ANE, we also measured dc resistivity and longitudinal Seebeck coefficient as a function of temperature and used these data to analyze our results. Our data analysis indicates that skew scattering is the main source of the ANE in these oxides.

Keywords: Anomalous Nernst effect; Spin-caloritronics; Manganites; Metal-insulator transition

*Corresponding author's email: phyrm@nus.edu.sg

## I. INTRODUCTION

Generation of electrical voltage from temperature difference ($\Delta T$) via Seebeck effect in non-magnetic semi-metallic alloys such as $Bi_2Te_3$, $Sb_2Te_3$ and semiconductors such as Si-Ge, SnSe etc. has drawn immense attraction because of harnessing waste heat from home appliances, automobile exhaust, industrial furnace and also low-grade heat from human body.[1,2] The Seebeck effect is caused by diffusion of charge carriers from hot to cold ends of a solid. The Seebeck voltage ($\Delta V_{xx} = S_{xx}\Delta T_x$) is measured along the direction of the temperature gradient where $S_{xx}$ is the longitudinal Seebeck coefficient. However, a voltage difference also appears in transverse direction to the temperature gradient if an external magnetic field is applied perpendicular to the direction of temperature gradient. This effect is known as the ordinary Nernst effect (NE) and it arises from deflection of charge carries by Lorentz force in a paramagnet ($\boldsymbol{E_{NE}} = Q\,(\boldsymbol{B}\times\nabla\boldsymbol{T})$ where $\boldsymbol{E_{NE}}$ is the electric field due to the Nernst effect, $\boldsymbol{B}$ is the magnetic flux density vector, $\nabla\boldsymbol{T}$ is the temperature gradient vector and $Q$ is Nernst coefficient). Hence, the Nernst voltage increases linearly with magnetic field in a paramagnet. In case of ferromagnets, spin-orbit interaction that selectively deflects up and down spin charges to opposite sides of the sample leads to anomalous Nernst voltage that often mimics the non-linear magnetic field dependence of magnetization rather than linear dependence on magnetic field as in a paramagnet. The Anomalous Nernst effect (ANE) is proportional to the cross product of magnetization vector ($\boldsymbol{M}$) and the $\Delta T$ ($\boldsymbol{E_{ANE}} = Q\mu_0\boldsymbol{M}\times\nabla\boldsymbol{T}$). The interplay between heat current and magnetism has opened a new arena of research known as spin-caloritronics whose aim is to harness spin degree of freedom of charge carriers to create Seebeck and Nernst voltages not only in conducting ferromagnets but also in insulating ferro/ferri/antiferromagnets.[3–7] Thermopile devices that can exploit ANE are simple in structure since they have lateral architecture and easy to fabricate compared to vertical pillar like architecture constructed from $n$ and $p$ junctions in conventional Seebeck devices.[8]

While ANE is currently being hotly pursued in nanometer thick magnetic films, there are not sufficient data available on bulk materials. Recently, large ANE were obtained in $Mn_3Sn$[9] and $Co_2MnGa$[10] single crystals. Among oxides, ANE was reported in single crystals of ferrimagnetic magnetite ($Fe_3O_4$),[11] ferromagnetic cobaltites ($La_{1-x}Sr_xCoO_3$, $x = 01$-$0.3$),[12] strontium ruthanate ($SrRuO_3$) and rare earth molybdenates ($R_2Mo_2O_7$, $R =$ Nd, Sm, $Gd_{0.5}Ca_{0.5}$).[13] All these work established that ANE and its electrical analogue, the anomalous Hall effect (AHE) are closely correlated. ANE and AHE have common contributions from asymmetric skew scattering,[11] side jump mechanism and the intrinsic Berry curvature.[14] For ANE, only the Berry curvature at Fermi level contributes whereas all the states contribute to AHE. Although manganites have been extensively investigated over last three decades for their colossal magnetoresistance and other physical properties, there is no report on ANE in single crystals of these oxides so far. ANE was reported in polycrystalline manganites ($La_{0.88}MnO_3$,[15] $La_{0.7}Ca_{0.3}MnO_3$,[16] $Ca_{0.88}Sm_{0.12}MnO_3$,[17] $Sm_{0.6}Sr_{0.4}MnO_3$[18]) by Suryanarayanan and collaborators. It is only during the past one year, spin Seebeck and ANE in manganite thin films were reported for a specific composition: $La_{0.7}Sr_{0.3}MnO_3$.[19,20]

In this report, we investigate the ANE in bulk polycrystalline samples of $La_{1-x}Na_xMnO_3$ (LNMO) with $x$ = 0.02, 0.05, 0.1 and 0.3. Since Na is a monovalent cation, it converts two of its neighboring $Mn^{3+}$ ions into $Mn^{4+}$ ions thus hole density is doubled compared to the substitution of divalent Sr or Ca cations for same $x$ value. Antiferromagnetic $LaMnO_3$ turns into a ferromagnetic metal at low temperatures with increasing Na content.[21] We report not only the temperature dependence of ANE in LNMO samples for a small field of $H$ = 1 kOe but also its field dependence up to $H$ = ±2 kOe at a fixed temperature of $T$ = 200 K. In addition to ANE, we also measured the temperature dependence of resistivity and longitudinal Seebeck effect in these materials. We show that ANE signal increases rapidly at the onset of ferromagnetic transition in all these samples and, in general, exhibit a peak just below the Currie temperature ($T_C$). Our data analysis indicates that ANE is dominated by skew scattering in this series of oxide.

## II. EXPERIMENTAL DETAILS

Polycrystalline $La_{1-x}Na_xMnO_3$ ($x$ = 0.02, 0.05, 0.1 and 0.3) were prepared by solid state reaction method using high purity $La_2O_3$, $MnO_2$ and $NaCO_3$ powders. $La_2O_3$ was preheated to 900ºC for 10 h prior to mixing. Stoichiometric amount of the well mixed powders was calcined at 1000ºC for 24 h followed by sintering at 1150ºC for 48 h in few steps of intermediate grinding. The samples were characterized by X-ray diffraction for purity and structure, by a vibrating sample magnetometer for magnetization, simultaneous electrical resistivity ($\rho$) and longitudinal thermopower ($S_{xx}$) measurement using a home-made sample stage for a physical property measuring system (PPMS). Ceramic pallets were cut in rectangular shapes (3 × 5 × 1 mm$^3$) for the ANE measurement which was performed from 35 K to 330 K and up to 2 kOe magnetic field using a home-built spin-caloritronic measurement setup. The schematic representation of the measurement configuration is given in Fig. 2(a). The sample is sandwiched between two copper blocks, separated electrically by a thin insulator. A temperature difference of $\varDelta T = 20 \pm 0.005$ K was applied along $z$ direction, magnetic field along $y$ direction, whereas the ANE voltages were picked up along $x$ direction. The sample temperature mentioned in the report corresponds to the average temperature, $T_{Sample} = (T_{Hot} + T_{Cold})/2$. The ANE voltages were measured at each temperature for both the positive and negative fields and calculated as $V_{ANE} = \big(V(+H) - V(-H)\big)/2$ to subtract the contribution of background signal.

## III. RESULTS AND DISCUSSION

Fig. 1(a) represents the temperature dependence of magnetization of polycrystalline La$_{1-x}$Na$_x$MnO$_3$ ($x$ = 0.02, 0.05, 0.1 and 0.3) samples measured in a field of 1 kOe. $M(T)$ increases rapidly in each sample when ferromagnetism sets in. The $T_C$ is determined from the inflection point of $dM/dT$ and it increases with increasing $x$ in La$_{1-x}$Na$_x$MnO$_3$ ($T_C$ = 251 K, 255 K, 276 K and 310 K for $x$ = 0.02, 0.05, 0.1 and 0.3). Fig. 1(b) shows field dependence of magnetization at 200 K ($< T_C$) for all the samples. Coercive field is small ($<$ 10 Oe) in all the samples at 200 K and $M$ at $H$ = 2 kOe increases from $x$ = 0.02 to 0.1 but decreases for $x$ = 0.3. This can be understood as follows: the density of $Mn^{4+}$ ions increases with Na content and exceeds that of $Mn^{3+}$ for $x$ = 0.3 $\left(La^{3+}_{0.7}Na^{+}_{0.3}Mn^{3+}_{0.4}Mn^{4+}_{0.6}O^{2-}_{3}\right)$. As the $Mn^{4+}$ content exceeds 50 % in $La_{1-x}A_xMnO_3$ (A = $Sr^{2+}$, $Ca^{2+}$), $Mn^{4+} - O - Mn^{4+}$ antiferromagnetic interaction tends to compete with $Mn^{3+} - O - Mn^{4+}$ ferromagnetic double exchange interaction and hence the saturation magnetization value is expected to decrease.[22]

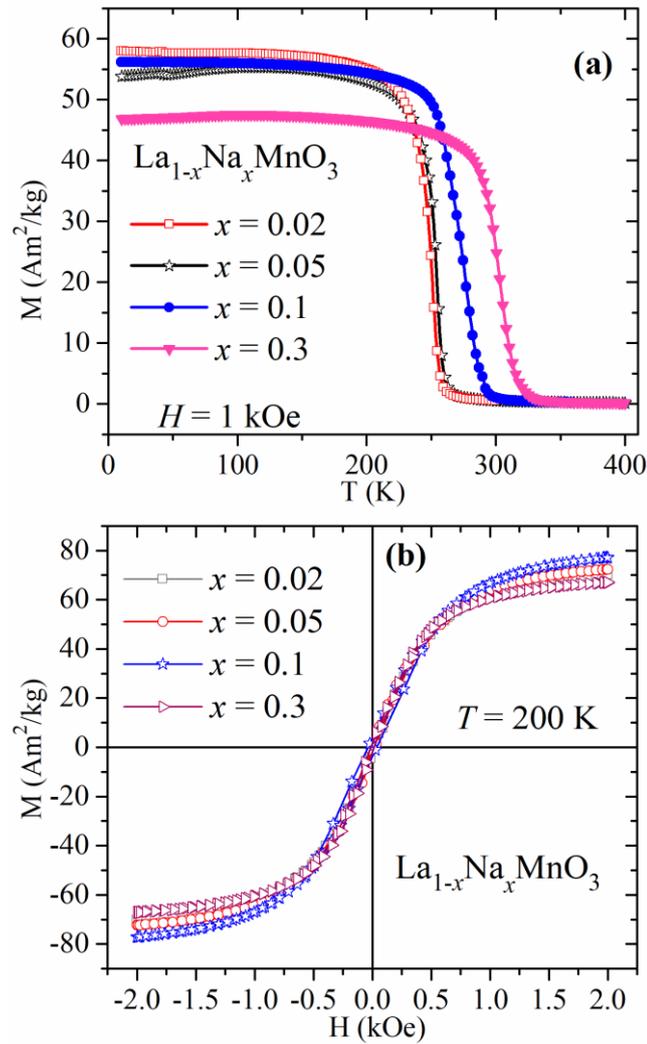

FIG. 1. (a) Temperature dependent magnetization measured under 1 kOe field and (b) field dependent magnetization at 200 K for $La_{1-x}Na_xMnO_3$ ($x$ = 0.02, 0.05, 0.1 and 0.3) samples.

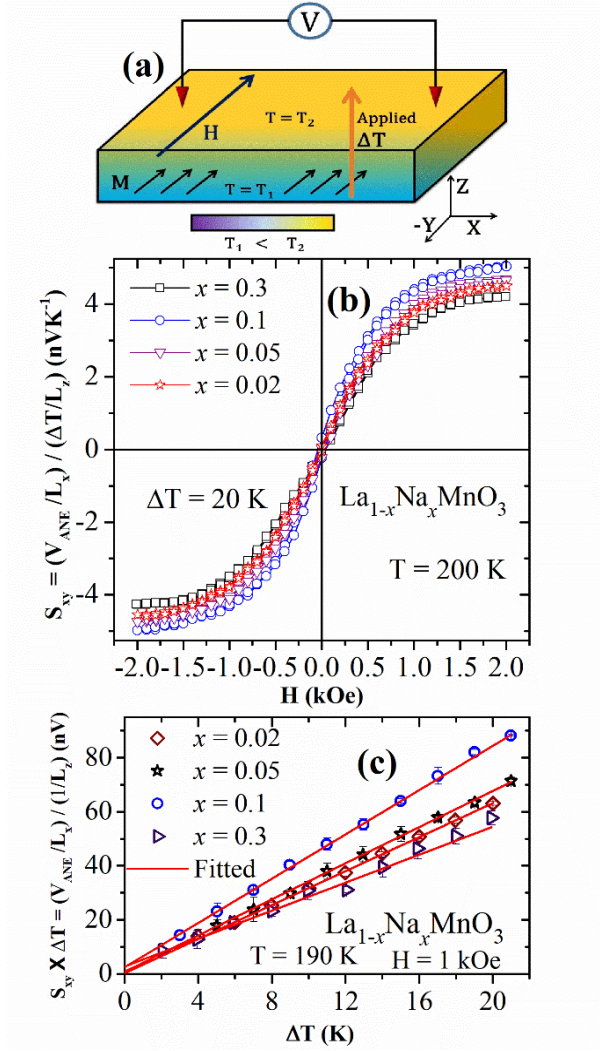

FIG. 2. (a) Schematic diagram of the ANE measurement configuration. (b) Magnetic field dependent transverse Seebeck coefficient ($S_{xy}$) for La$_{1-x}$Na$_x$MnO$_3$ ($x$ = 0.02, 0.05, 0.1 and 0.3) samples measured at 200 K under 20 K thermal gradient. (c) Normalized ANE voltages as a function of thermal gradient and their linear fitting.

Fig. 2(b) shows the magnetic field dependence of the transverse thermopower ($S_{xy}$) for La$_{1-x}$Na$_x$MnO$_3$ ($x$ = 0.02, 0.05, 0.1 and 0.3) samples measured at 200 K with $\Delta T$ = 20 K. The measured $S_{xy}$ has contributions from both the ordinary and anomalous Nernst signals (NE and ANE) and closely follows the field dependence of magnetization (Fig. 1(b) and 2(b)). Since NE is proportional to the applied magnetic field, it is obtained by linearly fitting the slowly varying portion of the $S_{xy}$ vs $H$ curves for H > 1 kOe. The contribution from NE is found to be ~ 0.25 nV K$^{-1}$ whereas the saturated value is ~ 4.5 nV K$^{-1}$. Thus, the major contribution to the measured signal is due to ANE. The saturation value of $S_{xy}$ is maximum for $x = 0.1$ sample. In Fig. 2(c) we plot the dependence of ANE signal for different $\Delta T$ values. $S_{xy}$ increases linearly with temperature difference ($S_{xy} \propto \Delta T$) and when extrapolated to $\Delta T \to 0$, the intercept passes through the origin, which implies that the measurements are free from artifact.

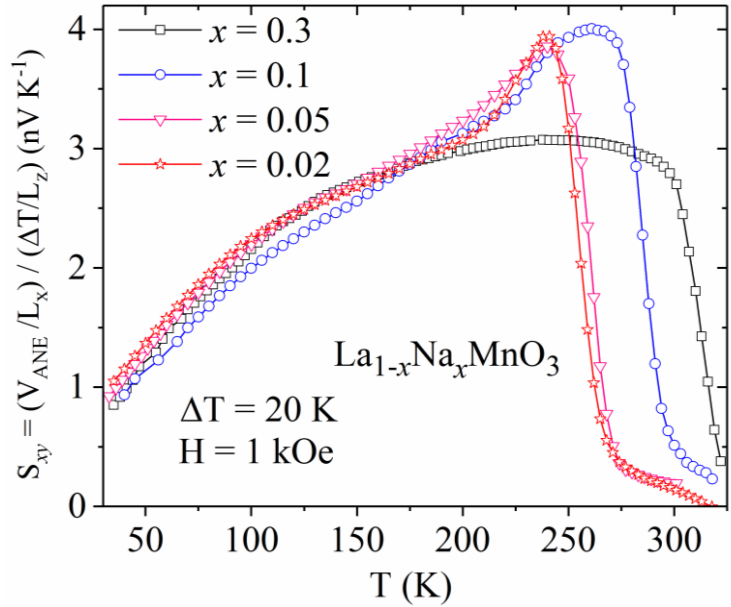

FIG. 3. Anomalous Nernst signal as a function of temperature for LNMO samples measured under 1 kOe magnetic field and 20 K thermal gradient.

Fig. 3 shows the temperature dependence of $S_{xy}$ for LNMO samples measured in a field of $H = 1$ kOe. For all the samples, $S_{xy}(T)$ decreases with the decreasing temperature below 175 K. As the population of thermally excited magnons are less at lower temperatures compared to higher temperatures,[23] transverse scattering probability of thermally excited carriers are minimized which results in decrease of $S_{xy}$ at low temperatures. As the temperature increases above 200 K, $S_{xy}$ increases and reaches a maximum value just below $T_C$ and decreases abruptly as the sample enters in to the paramagnetic state. Above $T_C$, $S_{xy}$ decreases monotonically with increasing the temperature. Since the samples are paramagnetic above $T_C$, the observed signal should originate only from NE. Since $T_C$ = 310 K of $x$ = 0.3 sample is around the high temperature limit (310 K) in our set up, we do not see completion of the ferro-paramagnetic transition in $S_{xy}(T)$. Compared to other samples, $S_{xy}(T)$ in $x$ = 0.3 shows a broad maximum below $T_C$. For a conductor with a single type of carrier, the Nernst coefficient $Q$ is dependent on temperature ($T$) and energy dependent relaxation rate of carrier as $Q = \frac{S_{xy}}{B} = \frac{\pi^2 k_B^2}{3m} T \left(\frac{d\tau}{dE}\right)_{E_F}$ where, $k_B$, $E$, $m$ and $\tau$ are respectively the Boltzmann constant, energy, effective mass of the carrier and energy dependent relaxation time of the carrier near Fermi level ($E_F$).[15] The increase in $S_{xy}$ with decreasing temperature in the paramagnetic state is due to decrease in hopping frequency of small polarons between neighboring Mn sites and hence to the increase in $\tau$. As ferromagnetism sets in at $T_C$, the conduction electron band width widens and the mobility of small polarons suddenly increases which results in peak of $S_{xy}(T)$. The maximum value of $Q$ value is 40 nVK$^{-1}$T$^{-1}$ for $x$= 0.02 and 30 nVK$^{-1}$T$^{-1}$ for $x$ = 0.3.

$S_{xx}$ of a conductor linearly dependents on temperature, the energy dependence of longitudinal conductivity ($\sigma_{xx}$) at the $E_F$ and it is given by the Mott's relation[24]

$$S_{xx} = \frac{\pi^2 k_B^2 T}{3e\sigma_{xx}} \left(\frac{\partial \sigma_{xx}}{\partial E}\right)_{E_F} \quad (1)$$

where $e$ is the electric charge. $S_{xy}$ is related to $S_{xx}$ through the equation[11]

$$S_{xy} = \rho\left(\alpha_{xy} - S_{xx}\sigma_{xy}\right) \quad (2)$$

where, $\sigma_{xy}$ is the transverse electrical conductivity and $\alpha_{xy}$ is the transverse thermoelectric conductivity (Peltier conductivity) which is related to the energy dependent $\sigma_{xy}$ through the Mott's relation[11]

$$\alpha_{xy} = \frac{\pi^2 k_B^2 T}{3e} \left(\frac{\partial \sigma_{xy}}{\partial E}\right)_{E_F} \quad (3)$$

In order to investigate the nature of interaction in the region $T \ll T_C$, we consider Eq. (1) and Eq. (2) which can be converted into the following form:[11]

$$S_{xy} = \rho_{xx}^{n-1}\left(\frac{\pi^2 k_B^2}{3|e|}T\lambda' - (n-1)\lambda S_{xx}\right) \quad (4)$$

where, $\rho_{xx}$, $\lambda$ and $\lambda'$ are respectively the longitudinal resistivity, spin-orbit coupling constant, first derivative of $\lambda$ and $n$ is a positive number which relates the transverse resistivity ($\rho_{xy}$) and $\rho_{xx}$ ($\rho_{xy} = \lambda\rho_{xx}^n$). For skew scattering mechanism, $n = 1$ whereas $n = 2$ for the intrinsic Berry curvature or side jump mechanism.[11] For side jump mechanism, the value of $S_{xy}$ depends linearly on $\rho_{xx}$ and $S_{xx}$. For skew scattering ($n = 1$) the second term in Eq. (4) vanishes and power of $\rho_{xx}$ becomes zero. In this case, ANE becomes independent of the bulk electrical transport and thermoelectric properties of a material. The value of $n$ is also dependent on the conduction mechanism and the magnitude of resistivity, For example, $Fe_{3-x}Zn_xO_4$ samples show metallic type of resistivity above the Verway transition but the value of resistivity is much higher than pure metals. In this regime, it is believed that charge carriers are polarons in a wide polaron band and $n$ was found to be 0.4.[11,25,26]

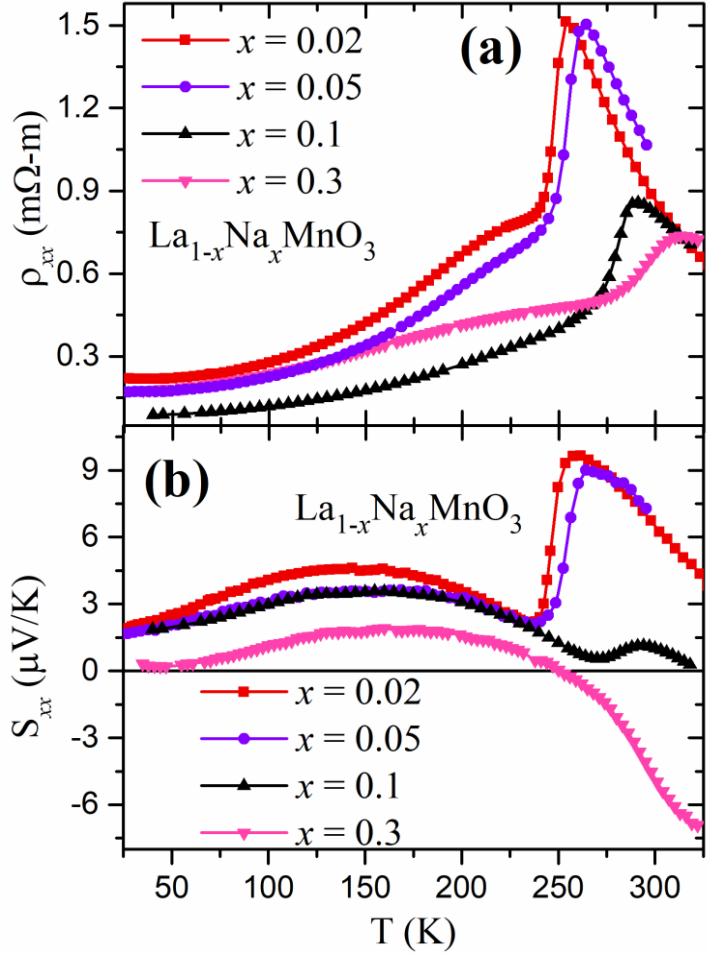

FIG. 4. (a) Temperature dependence of bulk resistivity for LNMO samples. (b) Longitudinal Seebeck coefficient ($S_{xx}$) as a function of temperature.

In Fig. 4(a) and 4(b), we show the temperature dependence of $\rho_{xx}$ and $S_{xx}$ in $La_{1-x}Na_xMnO_3$. These data are used to fit the $S_{xy}$ vs $T$ curves that will be discussed later in this paper. $\rho_{xx}(T)$ of all the samples show activated behavior in the paramagnetic state and metallic-like behavior in the ferromagnetic state. The transition from para to ferromagnetic state is accompanied by a peak in the resistivity (Fig. 4(a)). The peak value of $\rho$ decreases and the peak shifts to high temperature with increasing $x$. $S_{xx}$ versus $T$ also shows a positive peak at $T_C$ for $x$ = 0.02. 0.05, 0.1 and the magnitude of the peak decreases with increasing $x$ (Fig. 4(b)). The sign of $S_{xx}$ changes from positive for $x \leq 0.1$ to negative for $x$ = 0.3. The peak is absent for $x$ = 0.3. These trends of $S_{xx}$ are consistent with increasing $x$ trend found in $La_{1-x}Ca_xMnO_3$.[22]

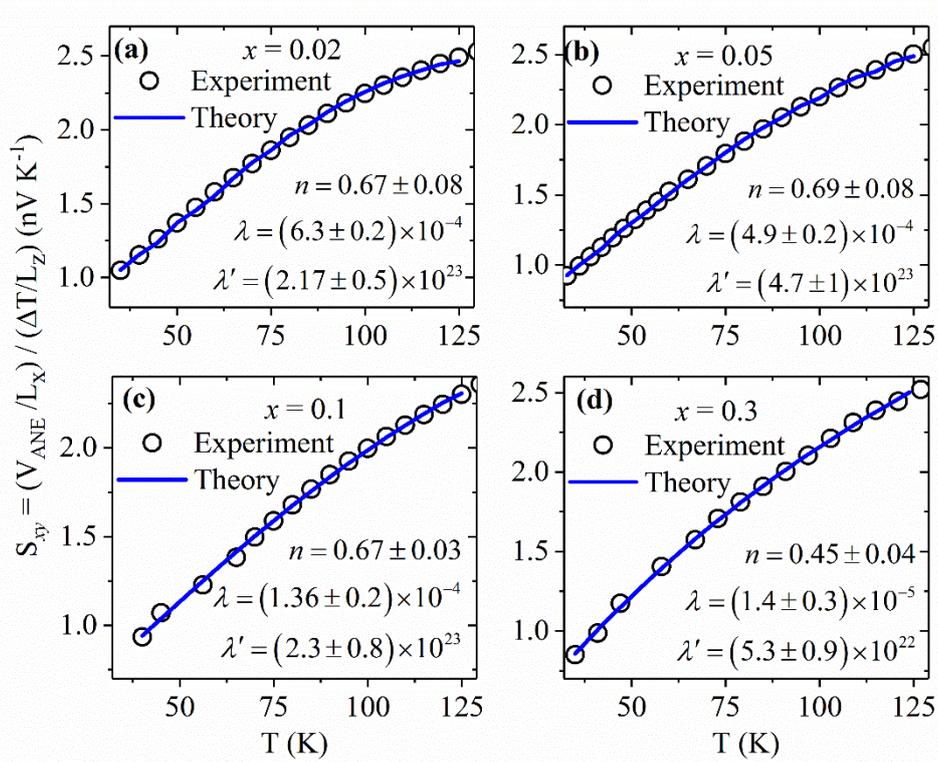

FIG. 5. Fitting of the temperature dependent transverse Seebeck coefficient $(S_{xy})$ data using Eq. (4) for $T \ll T_C$ of $La_{1-x}Na_xMnO_3$ samples.

Incorporating the results of Fig. 3 and 4, we have fitted $S_{xy}(T)$ in the region $T < 125$ K (well below $T_C$) by using Eq. (5) where $\lambda$, $\lambda'$, and $n$ were taken as fitting parameters. The fitted curves are shown in Fig. 5. Values of parameters with their error scale are mentioned in the respective figures. The value of $n$ is found to be in between 0.4 and 1 which means that the origin of ANE in these bulk Na doped LaMnO$_3$ is dominated by skew scattering.[11,26] The values of $\lambda$ in our LNMO samples are of similar order ($\lambda \sim 10^{-4}$) reported in Fe$_3$O$_4$ single crystal.[11] We note that $\lambda$ decreases with the increase in $x$ in our series. $\lambda$ holds a proportional relationship with Hall angle which decreases with decreasing $\rho$ in metallic materials.[27] This is consistent with the temperature dependent resistivity data (Fig. 4(a)) where $\rho$ decreases with increasing $x$.

## IV. CONCLUSION

In summary, we have systematically studied the anomalous Nernst effect (ANE) in the hole doped bulk La$_{1-x}$Na$_x$MnO$_3$ ($x$ = 0.02 0.05, 0.1 and 0.3) samples. The temperature dependence of ANE exhibits a peak just below $T_C$ in all the sample and the value of ANE increases up to $x$ = 0.1 and then decreases. It is found that ANE in Na doped manganites is dominated by the skew scattering mechanism. Since magnetic and electrical properties of manganites are dependent on conduction electron band width, hole content and structural distortions, it will be interesting to study ANE in other series of Mn oxides with different rare earth and alkaline earth combinations to enhance the value of ANE, and to check whether a common mechanism of ANE exists.

**ACKNOWLEDGEMENTS**

R.M. acknowledges the Ministry of Education (MOE), Singapore for supporting this work (Grant no. MOE 2014-T2- 114 and R144-000-147-112).

**Figure captions**

FIG. 6. (a) Temperature dependent magnetization measured under 1 kOe field and (b) field dependent magnetization at 200 K for $La_{1-x}Na_xMnO_3$ ($x$ = 0.02, 0.05, 0.1 and 0.3) samples.

FIG. 7. (a) Schematic diagram of the ANE measurement configuration. (b) Magnetic field dependent transverse Seebeck coefficient ($S_{xy}$) for $La_{1-x}Na_xMnO_3$ ($x$ = 0.02, 0.05, 0.1 and 0.3) samples measured at 200 K under 20 K thermal gradient. (c) Normalized ANE voltages as a function of thermal gradient and their linear fitting.

FIG. 8. Anomalous Nernst signal as a function of temperature for LNMO samples measured under 1 kOe magnetic field and 20 K thermal gradient.

FIG. 9. (a) Temperature dependence of bulk resistivity for LNMO samples. (b) Longitudinal Seebeck coefficient ($S_{xx}$) as a function of temperature.

FIG. 10. Fitting of the temperature dependent transverse Seebeck coefficient $(S_{xy})$ data using Eq. (4) for $T \ll T_C$ of $La_{1-x}Na_xMnO_3$ samples.